\documentclass[aps,prb,twocolumn,showpacs,superscriptaddress]{revtex4}

\usepackage{graphics}
\usepackage{amsmath}
\usepackage{amssymb}

\newcommand{\insertpic}[1]{\scalebox{0.31}{\rotatebox{-90}{\includegraphics{#1}}}} 
\newcommand{\insertpica}[1]{\scalebox{0.33}{\rotatebox{-90}{\includegraphics{#1}}}} 
\newcommand{\insertpictwo}[1]{\scalebox{0.45}{\rotatebox{-90}{\includegraphics{#1}}}} 
 
\newcommand{\bvec}[1]{{\mathbf #1}}

\begin{document}

\title{Heterogeneous slow dynamics 
in a two dimensional doped classical antiferromagnet}

\author{Malcolm P. Kennett}

\affiliation{TCM Group, Cavendish Laboratories, Cambridge University,
Madingley Rd, Cambridge, CB3 0HE, UK }

\author{Claudio Chamon}

\affiliation{Department of 
Physics, Boston University, Boston, Massachusetts 02215, USA}

\author{Leticia F. Cugliandolo}

\affiliation{Laboratoire de Physique Th\'{e}orique et Hautes Energies, 
Jussieu, 75252 Paris Cedex 05, France}
\affiliation{Laboratoire de Physique Th\'{e}orique de 
l'\'{E}cole Normale Sup\'{e}rieure,
Paris, France}

\date{\today}

\begin{abstract}
We introduce a lattice model for a classical doped two dimensional
antiferromagnet which has no quenched disorder, yet displays slow dynamics
similar to those observed in supercooled liquids. 
We calculate two-time spatial and spin
correlations via Monte Carlo simulations and find that for
sufficiently low temperatures, there is anomalous diffusion and
stretched-exponential relaxation of spin correlations. 
The relaxation times
associated with spin correlations and diffusion both diverge at low
temperatures in a sub-Arrhenius fashion if the fit is done over a
large temperature-window or an Arrhenius fashion if only low
temperatures are considered.
We find evidence of spatially heterogeneous dynamics, in which
vacancies created by changes in occupation facilitate spin flips on
neighbouring sites.  We find violations of the Stokes-Einstein
relation and Debye-Stokes-Einstein relation and show that  
the probability distributions of local spatial 
correlations indicate fast and slow populations of sites, and local spin
correlations indicate a wide distribution of relaxation times, similar to observations
 in other glassy systems with and without quenched
disorder.

\end{abstract}


\maketitle

\section{Introduction}
\label{sec:intro}
Doped two-dimensional antiferromagnets have attracted considerable
attention in the past two decades, due to their relevance as a model
to describe high temperature superconductivity in the
cuprates. \cite{Anderson,Bednorz} In addition to superconductivity,
there have been many other phases observed in the cuprates, and 
there has been particular recent interest in the spin glass observed
at low temperatures and intermediate
doping,\cite{Aharony,Birgeneau,Panagopoulos,Pana2} 
and the checkerboard ordered electronic state 
in the high-T$_c$ cuprate Bi$_2$Sr$_2$CaCu$_2$O$_{8+\delta}$ 
(Bi2212) and in hole-doped copper oxides.\cite{Check1,Hanaguri} 
Whilst quantum effects
are very important in the cuprates, it is also of interest to study
whether glassy phases and slow dynamics are present in simple
{\it classical} doped two dimensional antiferromagnets, and this is the
subject we address here.

In this paper we construct the simplest such model that we can find
that displays glassy dynamics - it has nearest neighbour interactions
only and the spins are on a square lattice, so there is no geometric
frustration.  The only source of frustration is that there are mobile
holes, and these mean that local relaxation to N\'eel order leads to
local frustration.  

Whilst the model we construct is principally that of an
antiferromagnet, it falls within the general class of models which
show slow dynamics without quenched disorder.  There has been intense
activity in the past few years investigating models with this
behaviour, especially kinetically constrained models, which have
trivial Hamiltonians but more complicated rules governing their
dynamics.\cite{FA,Jackle,Mauro,Munoz,Evans,Chandler,Ritort,Biroli,BG,Berthier} 
These models have been investigated
with the aim of shedding light on the local structure of slow dynamics
in glasses, dynamical heterogeneities.\cite{Harrowell,Sillescu,Ediger2} The
approach here is complementary to the work on kinetically constrained
models, as we study a model which has a more complicated Hamiltonian,
but relatively simple dynamics. The model we study also has some similarities to 
frustrated lattice gases\cite{Fierro,Mimo} and the hard square lattice gas.\cite{Ertel}

We find that this model displays slow dynamics at low temperatures,
both in the diffusion of particles and in the relaxation of spins.  
For the values of the parameters used (temperature, doping and ratio 
between the antiferromagnetic and nearest-neighbour repulsion) and the 
linear sizes analyzed the system reaches a stationary state after a 
transient. This state is characterized by antiferromagnetic or 
checkerboard order.  In the stationary state we
find unusual temperature dependence of the relaxation times for both
diffusion and spin relaxation, that can be fit by a sub-Arrhenius
temperature dependence over the entire temperature range that we study
(although there is Arrhenius
temperature dependence at low temperatures).  We
also find evidence of spatially heterogeneous dynamics that lead to a 
breakdown of the relation between diffusive and 
rotational motion dictated by the Debye-Stokes-Einstein law (where we consider the 
spin degree of freedom to model rotations). 
The distributions of local correlations measured at different time differences
are stationary, and hence trivially scale with the value of the global
correlation.\cite{Chamon3}
 The {\it form} of these distributions is interesting
in that they suggest that for spatial correlations there are two populations of sites, 
one with fast, and the other with
slow dynamics, and a wide range of spin relaxation times.

The paper is structured as follows.  In Sec.~\ref{sec:model} we
introduce the model that we study and the one-time and two-time
quantities that we calculate with Monte Carlo simulations.  In
Sec.~\ref{sec:results} we display our results for the phase diagram
and the correlation functions that we defined in Sec.~\ref{sec:model}.
We show evidence of time-scales that diverge at low temperatures, 
and find that both the diffusive motion and
spin flip dynamics are spatially heterogeneous.  We also investigate
the distributions of local two-time
quantities.  In Sec.~\ref{sec:SE}, we discuss how our results indicate a 
breakdown in the Debye-Stokes-Einstein relation.
Finally, in Sec.~\ref{sec:disc} we summarize and discuss
our results.

\section{Model}
\label{sec:model}
The Hamiltonian for the model is
\begin{equation}
{\mathcal H} = \sum_{ij}n_i n_j \left(V + J S_i S_j\right) , 
\label{eq:modeldef}
\end{equation}
where $n_i =$ 0, 1 is a density variable indicating occupation of a
site, and $S_i = \pm 1$ is an Ising spin attached to each particle
(i.e. when a particle moves site, so does its spin).  There is a
nearest neighbour repulsion with magnitude $V$, and a
nearest neighbour antiferromagnetic interaction with magnitude $J$.
The
particles have a hard-core constraint so that there is no double
occupancy of sites.
We are interested in the limit where $J/V < 1$ (in the limit where
$J/V > 1$, we find phase segregation of antiferromagnetic domains and
regions with no particles).  We study this model in the canonical 
ensemble by fixing the number of particles on a two dimensional
square lattice and define the hole concentration $x = 1 - N/L^2$,
where $N$ is the number of particles and $L$ is the lattice size.
This is essentially the classical, Ising limit of the $t-J-V$ model
studied by Kivelson and Emery and co-workers.\cite{KEL,Carlson}

We note that the model has some similarities to others that have been
introduced in the literature, in particular the frustrated lattice
gas,\cite{Fierro,Mimo} although in that case, the spin interactions
$J_{ij}$ are randomly drawn from a distribution, rather than being of
constant sign and magnitude.  In the limit that $J \to \infty$, this model has some
resemblance to the hard square lattice gas \cite{Ertel} (although the
spin degree of freedom means that the dynamics here have a different
flavour to that model).

When there is no doping, the ground state is an antiferromagnet
with N\'eel order.  This implies that the system can be divided into
two sublattices, each with ferromagnetic order, shifted by one lattice
spacing in the $x$ and $y$ directions with respect to each other.  In
this limit, the model can be made equivalent to a ferromagnet by
applying a spin-flip transformation on one sublattice.  However, once
there is doping and holes are free to move around, the transformation
is no longer applicable and the antiferromagnet and ferromagnet are
distinct.

We simulate this model using classical Monte Carlo simulations and
Metropolis dynamics.  The model has no intrinsic dynamics, hence we
impose the following dynamics. At each step we choose with 50 \%
probability either to attempt to move a particle to one of its neighbouring
sites or to flip its spin.\cite{footnote1}  
We only allow attempts to move a particle to an unoccupied site (with equal 
probabilities for the unoccupied sites).\cite{footnote2}
The attempt is then accepted with Boltzmann probability.  In each of the
particle move and spin flip steps, detailed balance is respected. 
The Monte Carlo time-unit is equal to the attempt of $N$ updates
(motion of particles or spin flips).

\subsubsection{Motion of holes}

Whilst naively it might be thought that an individual hole would be
localized due to leaving a string of overturned spins behind it, holes
are able to move diagonally in an antiferromagnetic background at no
energy cost by hopping around one and a half loops of a
plaquette.\cite{Trugman} The energy of the final state is the same as
that of the initial state, however this is an activated process with
activation energy $12 \, J$,\cite{Trugman}
as the hole flips spins into an
unfavourable energy configuration on its first lap of the plaquette
then lowers the energy to that of the original state as it goes round
for another half a lap.  Similarly, a pair of holes that are on
opposite corners of a plaquette can move along the diagonal that they
lie on, although the activation for this process can involve $V$ as
well as $J$.  At low doping, where holes interact very rarely, this
implies that there are a lot of states with the same energy, yet with large
barriers between them, which naturally leads to slow dynamics.

\subsection{Quantities Calculated}

We calculate quantities that should demonstrate spin and 
hole ordering and illustrate the dynamics of the spins and holes.  The
one-time quantities that indicate ordering are the
staggered magnetization that indicates N\'eel order of the spins
and a checkerboard order parameter that indicates ordering of
the holes.  As has been discussed extensively in the literature (see
e.g. Refs.~\onlinecite{CugKur,Bouchaud}), glassy dynamics are often 
seen more easily in 
two-time quantities if one-time quantities reach a
limit. In most of the cases we consider the one-time quantities 
saturate for high temperatures, but can saturate quite slowly as 
the temperature is decreased.   
We calculate the two-time correlations for spins and mean
square displacement for both high and low temperatures.

\subsubsection{One time quantities}
We calculate two one-time quantities.  These are the staggered magnetization

\begin{equation}
M_s = \left| \frac{1}{N}\sum_{x,y} (-)^{x+y}S_{x,y} \right| ,
\end{equation}
and a checkerboard order parameter

\begin{equation}
M_c = \left| \frac{1}{N}\sum_{x,y} (-)^{x+y}n_{x,y} \right| ,
\end{equation}
where $x$ and $y$ are the co-ordinates of a given lattice site and
$S_{x,y}$ and $n_{x,y}$ are the spin and occupation of the site $i =
(x,y)$ respectively.  We calculate these quantities as a function of
temperature to determine the phase diagram, which is shown in
Fig.~\ref{fig:phase}.

\subsubsection{Two time quantities}
\label{sec:quant_twotimes}

We calculate several two-time correlation functions, for both spatial
and spin correlations.  These are: the mean square
displacement
\begin{equation}
D(t,t_w) = \frac{1}{N} 
\sum_{\alpha}
\left|(\bvec{r}_\alpha(t) - \bvec{r}_\alpha(t_w))^2\right| ,
\end{equation}
where $\bvec{r}_\alpha$ is the position of the $\alpha^{\rm th}$
particle; a spin auto-correlation function
\begin{equation}
C(t,t_w) = \frac{1}{N} \sum_{\alpha} s_\alpha(t) s_\alpha(t_w) ,
\label{eq:cdef1}
\end{equation}
where $s_\alpha$ is the spin of the $\alpha^{\rm th}$ particle; and a
constant-site spin correlation function
\begin{equation}
C_{local}(t,t_w) = \frac{1}{N} \sum_i S_i(t)S_i(t_w),
\label{eq:cdef2}
\end{equation}
where $S_i$ is the spin at site $i$, and we note that $S_i$ may not be
the same spin at times $t$ and $t_w$ (although at low enough
temperatures they coincide unless $t$ and $t_w$ are very widely
spaced).  We expect the long time limits of $C(t,t_w)$ and $C_{local}(t,t_w)$
to have different behaviour if there is N\'{e}el order present in the system. As $t \to \infty$,
we expect $C(t,t_w) \to 0$, whereas $C_{local}(t,t_w) \to M_s(t)M_s(t_w)$.\cite{footnote}
The arguments for each of these limits are as follows.  If there is Neel order, then the
lattice can be divided into two sublattices, each of which has an opposite spin orientation.
Consider the spins on one sublattice, at long times $t$, diffusion will mean that half are
on the other sublattice, and will have their spin flipped relative to its orientation at time
$t_w$.  The contributions of the set of spins on the two sublattices to $C(t,t_w)$ will have
opposite signs and equal magnitude and hence we expect $C(t,t_w) \to 0$.  In contrast, if we
consider the same-site correlation, then if there is no flipping of the antiferromagnetic
order, $C_{local}(t,t_w)$ should decay to $M_s(t)M_s(t_w)$ reflecting the average value of the
staggered magnetization on the site at the two times.  We mainly focus on $C(t,t_w)$ but show that 
$C_{local}(t,t_w) - M_s(t)M_s(t_w)$ has qualitatively similar time dependence to $C(t,t_w)$.

\subsection{Parameters}
In all our runs we used random initial conditions. Most of 
the data presented are for $L=30$ though we also used larger
systems $L=40$, and 50, to determine the phase diagram.  We set $J/V = 0.2$
unless otherwise specified.
In our calculations we mainly use a waiting time of $t_w =
3200$ Monte Carlo steps (MCs) (we found that we had essentially identical results for a
shorter waiting time of $t_w = 320$ MCs, and a longer waiting time of $t_w =$ 32000 MCs).  We 
accumulated data for up to
$10^8$ MCs after the waiting time.

\section{Results}
\label{sec:results}

\subsection{One-time quantities}
\label{sec:results-onetime}

\begin{figure}[htb]
\insertpic{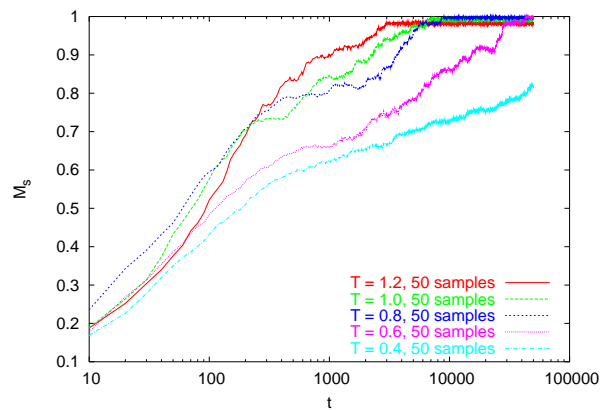}
\caption{(Color online) Staggered magnetization as a function of time averaged over
50 samples with $L = 30$ for $x = 0.1$, $J/V = 0.2$ and  temperatures $T/J =
1.2$, 1.0, 0.8, 0.6, and 0.4.}
\label{fig:saturate}
\end{figure}

\begin{figure}[htb]
\begin{center}
\insertpictwo{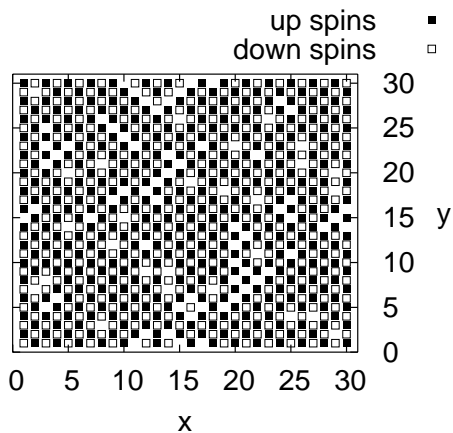}
\end{center}
\caption{Illustration of the state
of the system after 400 MCs in a sample with $L = 30$
for $x = 0.1$, $J/V = 0.2$, and $T/J = 0.1$. Up spins are indicated by black squares
and down spins are indicated by open squares.}
\label{fig:domain}
\end{figure}

We first display results for one-time quantities which demonstrate that
these saturate relatively quickly at high temperatures, but the saturation
can be quite slow as the temperature is lowered at low values of $x$.  The staggered
magnetization at five different
temperatures is shown as a function of time 
in Fig.~\ref{fig:saturate} for $x = 0.1$ and $J/V = 0.2$, averaged over
50 samples of size $L = 30$.  For this choice
of parameters, the configurations have
local N\'eel ordering at relatively short times, as shown in 
Fig.~\ref{fig:domain} for one sample at low temperature with one large
 domain.  However, there can
be domain walls that move slowly, and lead to a relatively slow saturation
of the staggered magnetization.  The two-time correlations defined in Eq.~(\ref{eq:cdef1})
do not appear to be
particularly sensitive to the presence of antiferromagnetic domain walls.

\begin{figure}[htb]
\insertpic{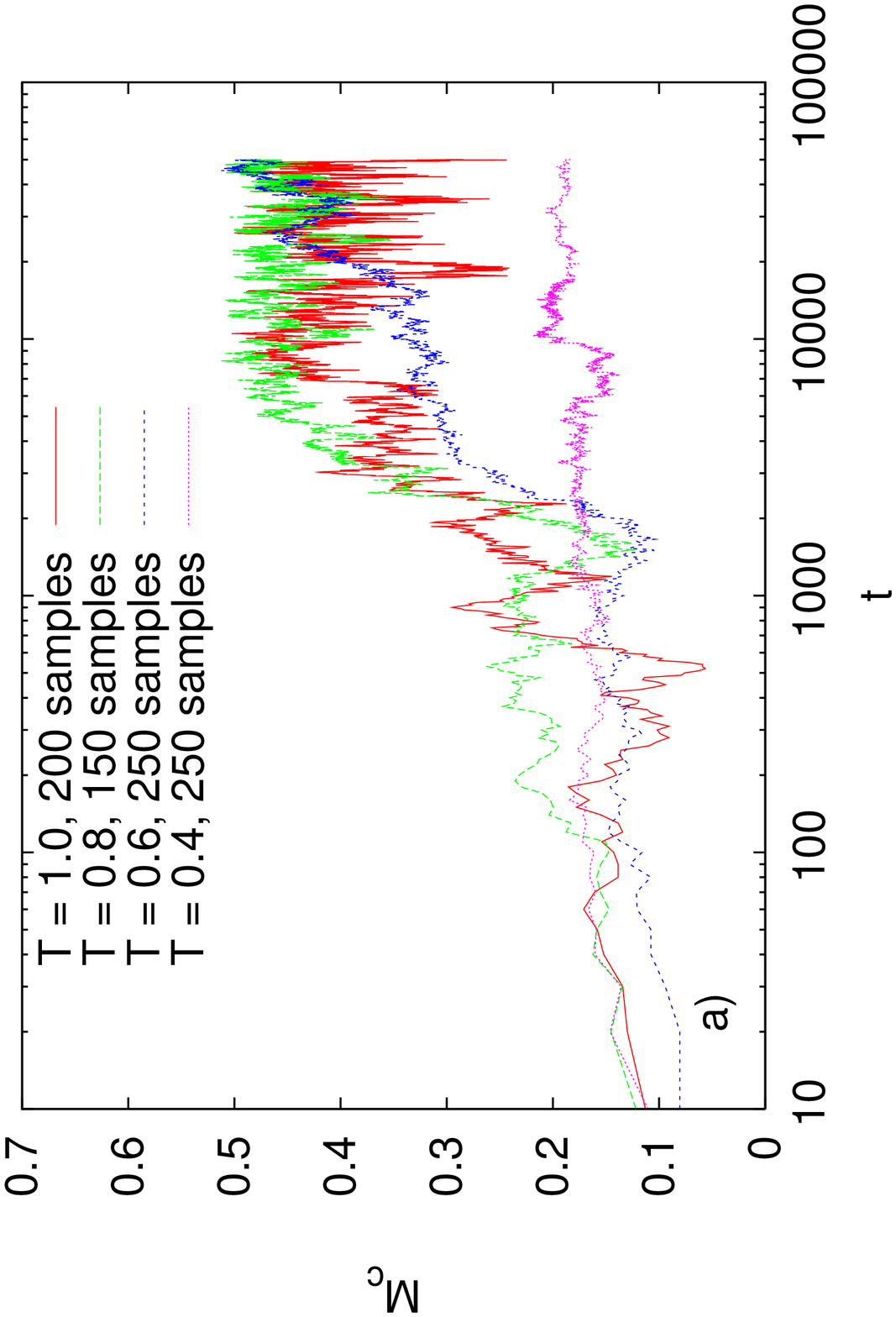}
\insertpic{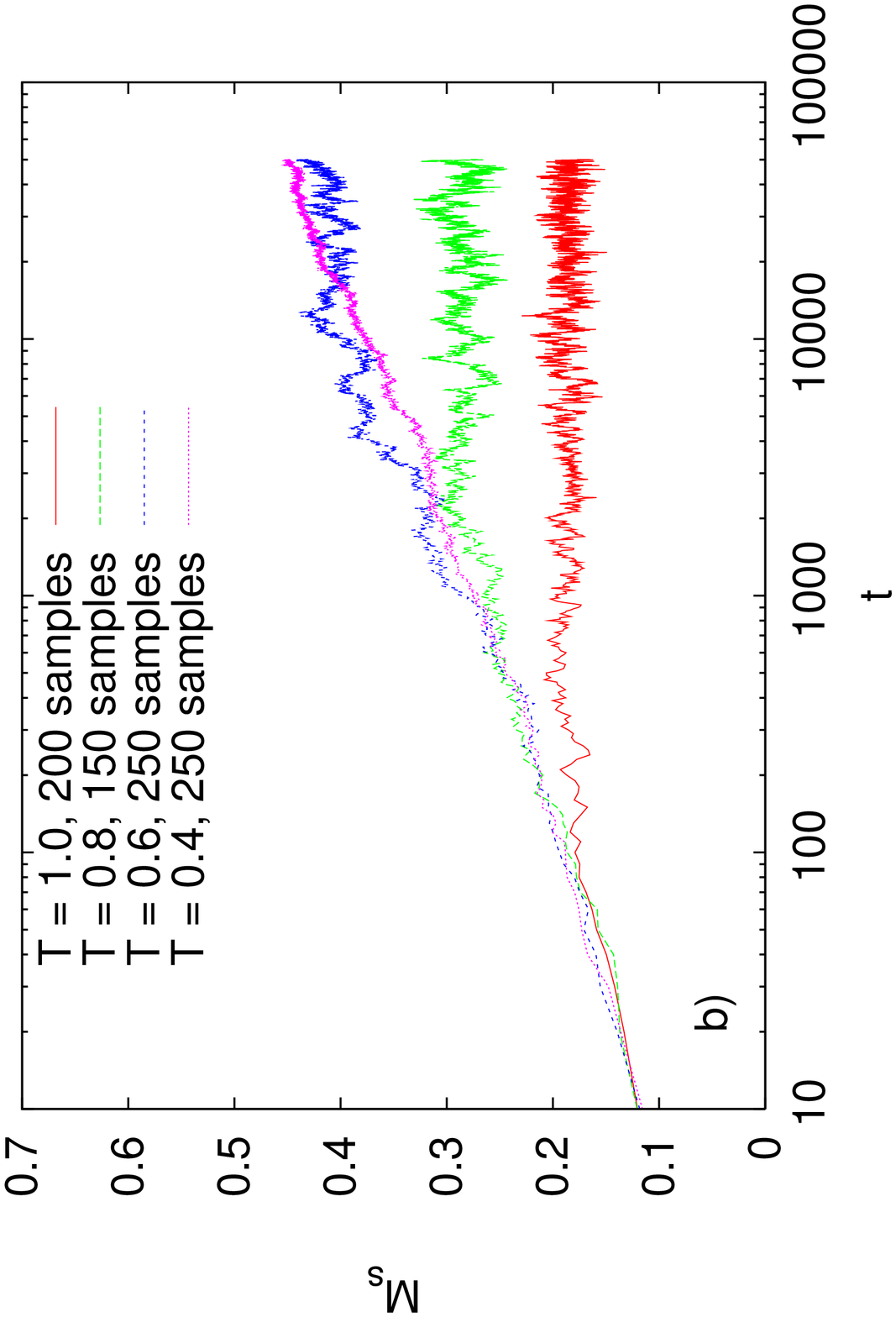}
\caption{(Color online) a) Checkerboard order parameter and b) staggered magnetization
as a function of time in Monte Carlo steps averaged at least 100
samples with $L = 30$ for $x = 0.35$, $T/J = 0.4$, 0.6, 0.8, and 1.0, and
$J/V = 0.2$.}
\label{fig:saturate-check}
\end{figure}

\begin{figure}[htb]
\insertpictwo{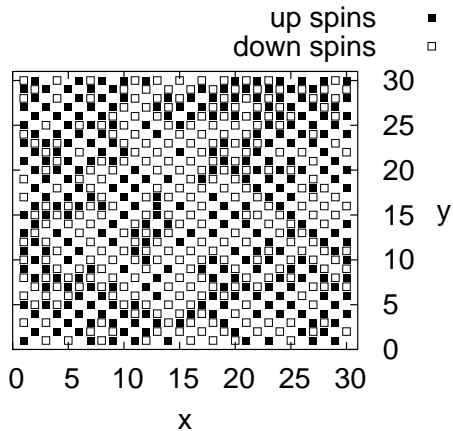}
\caption{A configuration with checkerboard hole-ordering, and N\'{e}el ordering for
$x = 0.35$, $L = 30$, $J/V = 0.2$, and $T/J = 0.1$ after 25600 MCs.}
\label{fig:check}
\end{figure}

At larger values of $x$, there is checkerboard order in addition to 
N\'{e}el order, and in  
Fig.~\ref{fig:saturate-check} we show the time-evolution 
of the checkerboard order parameter $M_c$ and staggered magnetization 
$M_s$ averaged over many samples for  $L = 30$ with $x = 0.35$ at four different
temperatures.  Figure~\ref{fig:check} shows
the configuration reached by a low temperature sample at the 
same doping level after 25600 MCs. 
The results in the two figures appear to be consistent with the
coexistence of antiferromagnetic and checkerboard order at low temperatures (the
non-zero value of $M_s$ at $T \gtrsim 0.8$ in Fig.~\ref{fig:saturate-check} b) is a
finite size effect), and 
Fig.~\ref{fig:check} appears consistent with the possibility of
phase separation.  We note that the time-scale for the onset of checkerboard
order grows as temperature is decreased so that for $T = 0.4$, the checkerboard order
has not saturated in the time-window considered, even though at higher temperatures 
in Fig.~\ref{fig:saturate-check} a)
the value of the order parameter saturates at around $M_c \simeq 0.47$.  

\begin{figure}[htb]
\insertpic{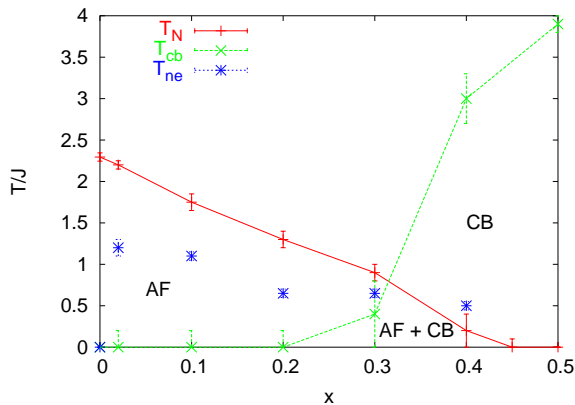}
\caption{(Color online) Phase diagram, showing
N\'{e}el temperature $T_N$, checkerboard ordering temperature $T_{cb}$, and
onset temperature of
non-exponential spin relaxation, $T_{ne}$ as a function of $x$ and $T/J$,
for $J/V = 0.2$.  The antiferromagnetic (AF) and checkerboard (CB) regions are marked,
as is the region of apparent coexistence of antiferromagnetic and checkerboard
order (AF + CB).}
\label{fig:phase}
\end{figure}

We show the phase diagram in Fig.~\ref{fig:phase} and indicate
the regions in which there is N\'{e}el ordering and checkerboard
ordering.  The boundary of the region of N\'{e}el order, the N\'{e}el
temperature, $T_N$, is found by calculating 
$M_s$ at $L=30$, 40, and 50 and extrapolating to where $M_s$ vanishes.
 We also find that for $x = 0$,
$T_N = 2.29 \pm 0.05 \, J$, in agreement with the exact value of $T_N =
2.27 J$ for the two-dimensional Ising antiferromagnet.
We use the same procedure with $M_c$ to calculate the checkerboard
ordering temperature $T_{cb}$.  The third temperature shown on the phase
diagram is $T_{ne}$, which is the temperature below which we see stretched
exponential spin relaxations, and is discussed further in Sec.~\ref{sec:two-time}.
From Fig.~\ref{fig:phase} it is clear
that N\'eel order disappears as $x$ is increased towards $0.5$, and
checkerboard order grows for $x \gtrsim 0.2$.  For immobile holes, 
one expects the antiferromagnetism to disappear at the percolation 
threshold.  
 The percolation
threshold for bond dilution in two dimensions is $x_c = 0.5$,
\cite{Newman} and for site dilution it is $x_c =
0.41$.\cite{Stauffer,Orbach} 
For any one snapshot of the
spin-hole configurations, there is no percolating cluster for $x >
0.41$ in the thermodynamic limit, and hence we would expect that $x_c = 0.41$.  
Whilst we have not investigated this question in great detail, our
results are consistent with antiferromagnetism vanishing at the site 
dilution threshold.  The slow dynamics typically
occur at temperatures lower than the N\'eel temperature, although we note
that $T_{ne} > T_N$ for $x = 0.4$, in the region where $T_{cb} > T_{ne} > T_N$.

Changing the ratio of $J/V$ whilst keeping it less than 1 appears to
have little effect on the N\'eel order, but the checkerboard
transition temperature decreases rapidly as $J/V$ increases.  For
instance, if $J/V = 0.5$ then we find that $T_{cb}/J$ is reduced to
1.3 at $x = 0.5$, compared to $T_{cb}/J = 3.8$ for $J/V = 0.2$, as
shown in Fig.~\ref{fig:phase}.  This appears to indicate that $T_{cb}
\sim V$ at low $J/V$, whilst $T_N \sim J$; we have not explored the
dependence of $T_{ne}$ on $J$ and $V$. 

\subsection{Two-time quantities}
\label{sec:two-time}

We show the spatial and spin correlation functions as a function of $t
- t_w$ at several different temperatures below.  All data shown here
for two-time correlations was taken in $L = 30$ systems -- the data
shown is only for one thermal history, although it was checked that
changing thermal history does not quantitatively change the results or fits.  We note
that there is some waiting time dependence of the results at low
temperatures, although this is generally only for very short waiting
times ($t_w < 320$ MCs).  Hence for the data we show here, we actually
find that the two-time quantities depend only on time differences for
the waiting times we use, so $D(t,t_w) \simeq D(t - t_w)$ and
$C(t,t_w) \simeq C(t - t_w)$.

Our analysis of the data follows the following scheme.  We attempt to fit 
\begin{equation}
D(t,t_w) = ((t-t_w)/\tau_r)^\alpha ,
\end{equation}
 and 
\begin{eqnarray}
C(t,t_w) & = & A_1 \, e^{-((t-t_w)/\tau_s)^\beta},  \\
C_{local}(t,t_w) & = & A_2 \, e^{-((t-t_w)/\tau_s)^\beta} + B, 
\end{eqnarray} 
where $\tau_r$ is the relaxation time for diffusion and $\tau_s$ is
the relaxation time for spin correlations. 
Apart from very short times and very
long times (where the system size becomes important for $D$, and spin
correlations die out for $C$), these forms work very well (note that
$A_1 \to 1$, $A_2 \to 0$, and $B \to 1$ 
at low temperatures and low doping).  We use the fits to the
two-time spin correlation to fit $T_{ne}$, the highest temperature at
which one observes non-exponential relaxation, {\it i.e.} $\beta < 1$. 
This temperature scale is not associated with a change in any order parameter 
that we measure, but allows us to determine the boundary of the region
in the phase diagram where slow dynamics are observed.  However, we do note that
the equilibration of one-time quantities at $T < T_{ne}$ is generally much slower than
for $T > T_{ne}$.

\begin{figure}[htb]
\insertpica{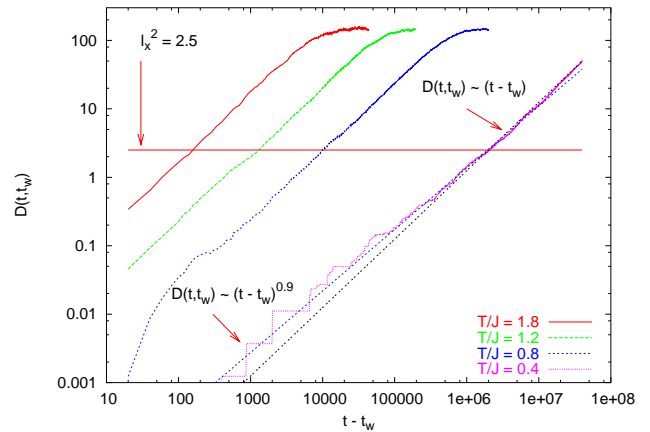}
\caption{(Color online) Spatial correlations for $x = 0.1$, $L = 30$, and $J/V = 0.2$ for
temperatures $T/J = 0.4$, 0.8, 1.2 and 1.8.  Note that at the lowest
temperature the slope of the curve changes, indicating anomalous
diffusion.  We also indicate $l_x^2 = 1/4x = 2.5$ and the fits to
$D(t,t_w)$ at $T = 0.4$ for $D(t,t_w) < l_x^2$ and $D(t,t_w) > l_x^2$.
Note also that the saturation at long times and high temperatures is a
finite size effect.}
\label{fig:ddata}
\end{figure}

In the figures below, we only show data from $x = 0.1$ unless indicated, but it is
representative of the behaviour seen at other levels of doping, where
we observe similar qualitative behaviour, with only quantitative
differences.  In Fig.~\ref{fig:ddata} we show the mean square displacement 
as a function of time
for $x = 0.1$ for $J/V = 0.2$.

\begin{figure}[htb]
\insertpica{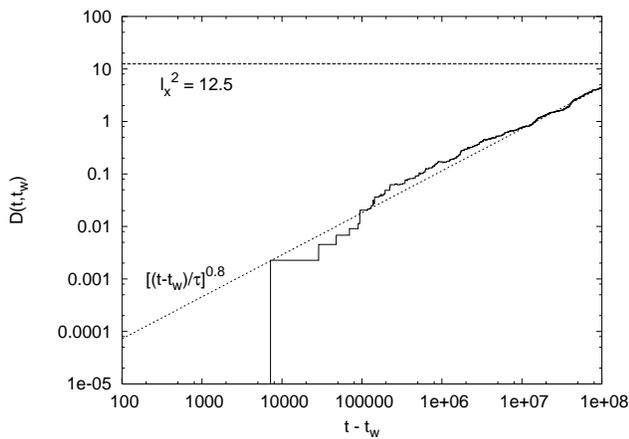}
\caption{Spatial correlations for $x = 0.02$, $L = 30$, and $J/V = 0.2$ for
temperatures $T/J = 0.4$. Anomalous diffusion is clearly evident.  We
also indicate $l_x^2 = 1/4x = 12.5$ and a fit to $D(t,t_w)$ of the
form $((t-t_w)/\tau)^{0.8}$.}
\label{fig:anomdiff}
\end{figure}

Note that at the lowest temperature in Fig.~\ref{fig:ddata} there is a
signature of anomalous diffusion (i.e.  $\alpha < 1$).  This is
present for mean square displacements that are less than $l_x^2 =
1/4x$, which is the square of the lengthscale equal to half the
average distance between holes. We show a more striking example
of anomalous diffusion in Fig.~\ref{fig:anomdiff} for $x = 0.02$,
where the lengthscale $l_x^2$ is much larger than in
Fig.~\ref{fig:ddata}.  In general we only see anomalous
diffusion at temperatures $T < T_{ne}$ and $D(t,t_w) < l_x^2$, however
the temperature where we first observe anomalous diffusion does not
appear to be related to $T_{ne}$, which is when the spin auto-correlations
show non-exponential relaxation.  We note that anomalous diffusion
has been observed in several experimental glassy
systems,\cite{Weeks,Pouliquen} and can be understood as ``caged''
motion on short lengthscales and timescales, with usual diffusion
taking over at longer timescales and lengthscales when particles have
escaped from their cages.  Here, the lengthscale for caging is set by
the mean distance between vacancies, $l_x$.

\begin{figure}[htb]
\insertpic{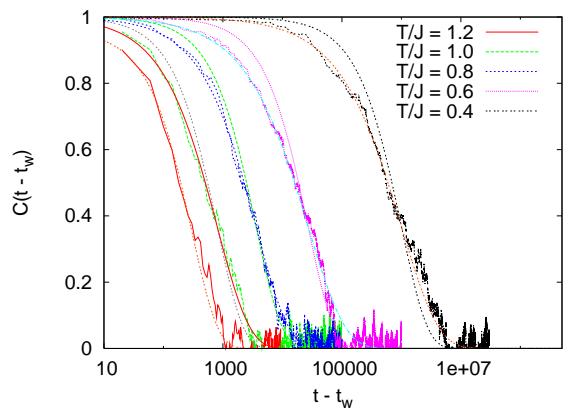}
\caption{(Color online) Spin auto-correlations for $x = 0.1$, $L = 30$, and $J/V = 0.2$, with
$T/J = 1.2$, 1.0, 0.8, 0.6, and 0.4. For $T/J \leq 1.0$ we display
both exponential and stretched exponential fits to the correlations --
in each of these cases, the stretched exponential is the better fit.}
\label{fig:cdata}
\end{figure}

\begin{figure}[htb]
\insertpic{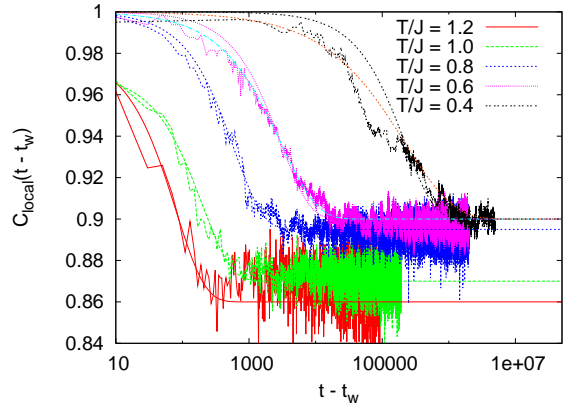}
\caption{(Color online) Spin same-site correlations for $x = 0.1$, $L = 30$, and $J/V = 0.2$,
with $T/J = 1.2$, 1.0, 0.8, 0.6, and 0.4. For $T/J \leq 0.6$ we
display both exponential and stretched exponential fits to the
correlations -- in each of these cases, the stretched exponential is
the better fit.}
\label{fig:cdata2}
\end{figure}

In Fig.~\ref{fig:cdata} we show the spin correlations for $x = 0.1$
and $J/V = 0.2$ for several different temperatures.  
We note that the timescale
for $C(t,t_w) \to 0$ is of the same order as the timescale for $D(t,t_w) 
\sim 1$, corresponding to the average particle having moved one lattice
spacing.  The argument explaining why $C(t,t_w) \to 0$ as $t \to \infty$
in Sec.~\ref{sec:quant_twotimes} gives a natural explanation of why these
timescales should be of the same order, and indicates the importance of
the interplay between spin and diffusion dynamics, as a particle will generally be 
forced to flip its spin as it diffuses.
For $T/J \leq 1.0$ we show both exponential and
stretched exponential fits to $C(t,t_w)$.  The stretched exponentials
are clearly better fits at low temperatures. 
We also performed calculations of
$C_{local}(t,t_w)$ for the same parameters as in
Fig.~\ref{fig:cdata}, and these correlations are shown in
Fig.~\ref{fig:cdata2}.  Whilst $C_{local}(t,t_w)$ decays to a different limit than
$C(t,t_w)$ it also displays stretched-exponential spin relaxations at 
low temperatures, although at a slightly lower temperature than the $T_{ne}$
defined by $C(t,t_w)$.

\begin{figure}[htb]
\insertpic{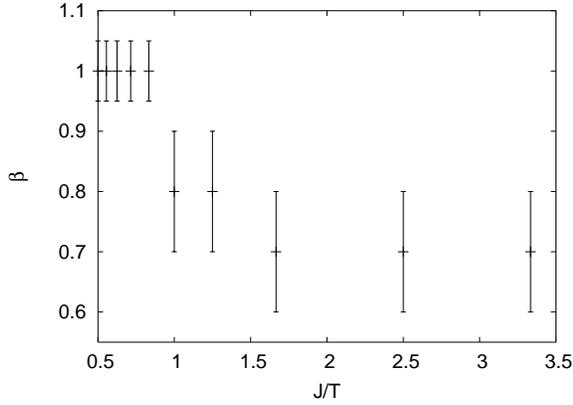}
\caption{Stretched exponential $\beta$ parameter
as a function of
temperature for $x = 0.1$, $L = 30$, and $J/V = 0.2$.}
\label{fig:beta}
\end{figure}

In Fig.~\ref{fig:beta} we show the stretched exponential
$\beta$ parameter as a function of temperature for the spin auto-correlations
shown in Fig.~\ref{fig:cdata}, which clearly indicates that the spin correlations
do not decay exponentially in time at low temperatures.
We also performed simulations for $x = 0$, in which we found no decay
of the spin correlations, indicating that the 
timescales observed here arise
solely due to doping.

\begin{figure}[htb]
\insertpic{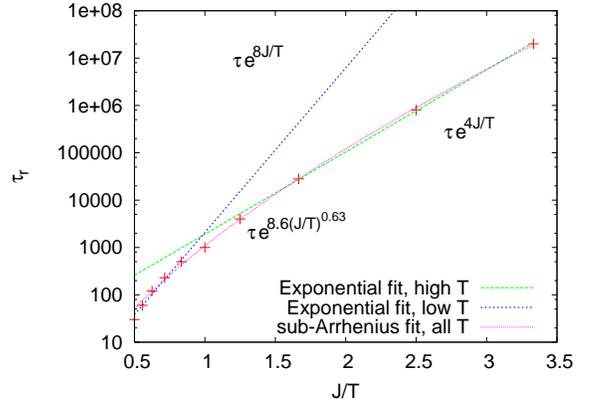}
\caption{(Color online) Fit to diffusion relaxation times determined from
spatial correlation functions as a function of
temperature at $x = 0.1$ and $J/V = 0.2$.}
\label{fig:taudata1}
\end{figure}

\begin{figure}[htb]
\insertpic{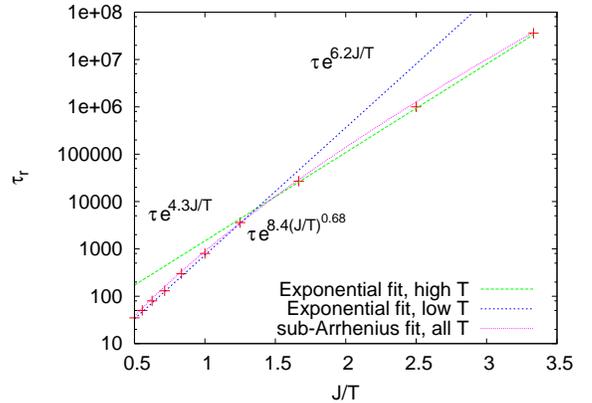}
\caption{(Color online) Fits to spin relaxation times determined from spin
correlation functions as a function of
temperature at $x = 0.1$, $L = 30$ and $J/V = 0.2$.}
\label{fig:taudata2}
\end{figure}

We extract values of $\tau_r$ and $\tau_s$ from fitting $D(t,t_w)$ and
$C(t,t_w)$ as described above, and use these to define relaxation
time-scales that we then fit as a function of temperature:
\begin{equation}
\tau_{r,s} =  \tau_0^{r,s} + \gamma_{r,s} 
\exp^{\left(\frac{E_0^{r,s}}{T}\right)^{a_{r,s}}},
\label{eq:taufit}
\end{equation}
where $\gamma_{r,s}$, $\tau_0^{r,s}$, $E_0^{r,s}$ and $a_{r,s}$ are
fitting parameters.  We also tried fitting our data to a Vogel-Fulcher
form, but this did not lead to good fits.  We show fits of the form in
Eq.~(\ref{eq:taufit})
for $\tau_r$ and $\tau_s$ determined for $x = 0.1$, $J/V = 0.2$ in
Figs.~\ref{fig:taudata1} and \ref{fig:taudata2}.  
In Table~\ref{tab:table1} we show values of $E_0$ and $a$ that we have
extracted for $\tau_r$ and $\tau_s$ in fits of the form of
Eq.~(\ref{eq:taufit}) for different $x$.  Those relating to $\tau_r$ are $E_0^r$ and
$a_r$ and those relating to $\tau_s$ are $E_0^s$ and $a_s$.

\begin{table}
\begin{tabular}{ccccc} \hline\hline
$x$  & $E_0^r$ & $a_r$ & $E_0^s$ & $a_s$ \\
\hline\hline
0.02 &  10.3 & 0.58 & 10.9 & 0.62 \\
0.1 &  8.6 & 0.63 & 8.4 & 0.68 \\
0.2 & 7.5 & 0.68 & 7.7 & 0.71 \\
0.3 & 6.9 & 0.76 & 7.1 & 0.77 \\
0.4 & 6.4 & 0.77 & 6.7 & 0.82  \\ \hline\hline
\end{tabular}
\caption{$E_0$ and $a$ data from fits to $\tau_r$ and $\tau_s$ made
using Eq.~(\ref{eq:taufit})}
\label{tab:table1}
\end{table}

The slowing down with temperature is
similar to that seen in glass formers.  However, there is a difference
-- in glass formers, the exponent $a$ is equal to 1 for strong glass
formers (Arrhenius) and 2 for fragile glass formers
(super-Arrhenius)\cite{Angell} whereas the relaxation times found here
consistently have $a<1$, i.e. sub-Arrhenius behaviour if one tries to
fit over the entire temperature range in which there is a 6 order of
magnitude change in the relaxation time.  We did find however, that if
we restricted the temperature range, then there is Arrhenius temperature
dependence of the relaxation times at low temperatures, as is also indicated in
Figs.~\ref{fig:taudata1} and \ref{fig:taudata2}.  This suggests this model
has strong-glass like behaviour at low temperatures. 

 The diffusion
relaxation time, $\tau_r$, and spin relaxation time $\tau_s$, display purely Arrhenius behaviour at low
temperatures ($T < T_{ne}$), and it is possible 
to get a good fit to both $\tau_r$ and $\tau_s$ with an Arrhenius form at high
temperatures $T > T_{ne}$,
but neither fit is good over the entire temperature range, as is visible
in Figs.~\ref{fig:taudata1} and \ref{fig:taudata2}.

\subsection{Spin configurations}
\label{sec:spinconfig}

\begin{figure}[htb]
\insertpictwo{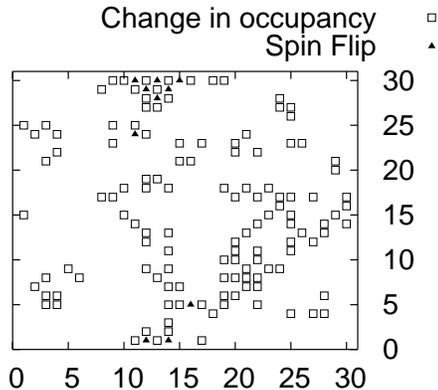}
\caption{Illustration of the sites at which there has been a change in occupancy or a spin flip
between times 400 MCs and 25600 MCs for $x = 0.1$, $T/J = 0.1$, $J/V = 0.2$, and $L = 30$.}
\label{fig:snap1}
\end{figure}

\begin{figure}[htb]
\insertpictwo{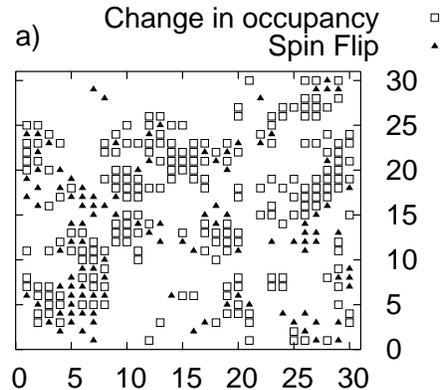} \\
\insertpictwo{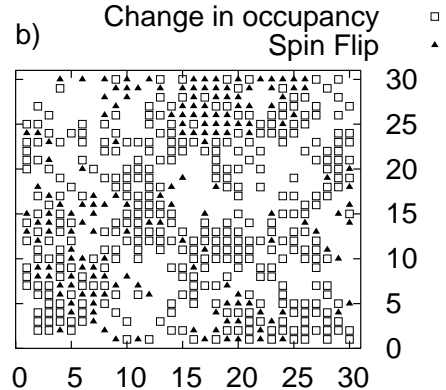}
\caption{The two figures illustrate the sites at which there has been
a change in occupancy or a spin flip between times 400 MCs and 3200
MCs for a) and between 400 MCs and 25600 MCs for b).  The parameters
are $x = 0.3$, $T/J = 0.1$, $J/V = 0.2$, and $L = 30$.}
\label{fig:snap2}
\end{figure}

In Figs.~\ref{fig:snap1} and \ref{fig:snap2} we show how changes in
the configurations of spins and occupancy occur in the region where
slow dynamics dominate.  The doping levels used are $x = 0.1$ (Fig.~\ref{fig:snap1})
and $x = 0.3$ (Fig.~\ref{fig:snap2}), and both figures are for low
temperatures ($T/J = 0.1$).
It is obvious from these plots that the
dynamics is quite spatially heterogeneous at this temperature.  
Spin flips occur in the
vicinity of holes and are facilitated by changes in occupation on
adjacent sites.  There are also regions in which there are no changes in
spin or occupation during the time-window employed. 
This implies that the dynamics are a
combination of vacancy motion and spin flips rather than being associated
with domain wall motion.  Note that the figures show where there has
been a net change in occupation or a net spin flip, rather than
whether this is the only change that has occurred between $t$ and $t_w$.

\subsection{Distributions of correlations}

\begin{figure}[htb]
\insertpic{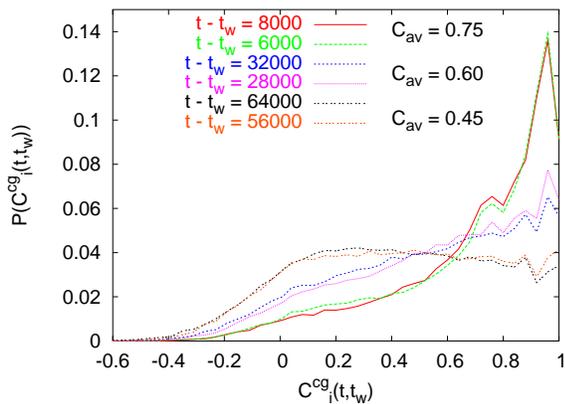}
\caption{(Color online) Distribution of coarse-grained local spin auto-correlations
for various values of $t - t_w$  with
$x = 0.1$, $T/J = 0.1$, $J/V = 0.2$, averaged over 20 samples of size $L = 50$.  We show the
average value of the global spin correlation, $C_{\rm av}$ for each pair of samples.}
\label{fig:dist2}
\end{figure}

In this section we show distributions of local spin
correlations and displacements.\cite{Chamon}  To obtain these 
distributions, we coarse-grain the correlations that give 
$D(t,t_w)$ and $C(t,t_w)$ when averaged over all sites, over a small
region about each site.  More precisely, we define
the coarse-grained local square displacement $D^{\rm cg}_i(t,t_w)$ and coarse-grained local
spin correlation $C^{\rm cg}_i(t,t_w)$ as

\begin{eqnarray}
D^{\rm cg}_i(t,t_w) & = & \frac{1}{A}\sum_{\alpha \in A_i(t)} 
\left(\bvec{r}_\alpha(t) - \bvec{r}_\alpha(t_w)\right)^2 , \\
C^{\rm cg}_i(t,t_w) & = &  \frac{1}{A}\sum_{\alpha \in A_i(t)} s_\alpha(t) s_\alpha(t_w),
\end{eqnarray}
where $A$ is the area of the box $A_i(t)$, which is the box centered on site $i$, 
containing the set of particles $\{\alpha\}$ at time $t$. In all of the histograms shown below, we
choose $A = 3^2 = 9$.  In
Fig.~\ref{fig:dist1} we show the distribution of coarse-grained local 
square displacements for various values of $t - t_w$.   We find
that for $t$ and $t_w$ with the same global mean square
displacement, i.e. $D(t,t_w)$, the
distributions collapse on each other.  In Fig.~\ref{fig:dist2}, we plot
similar distributions for coarse-grained local spin auto-correlations and
also see a clear scaling when the global correlation is the
same. We note that for time separations such
that $D(t,t_w)$ is the same,
$C(t,t_w)$ also coincides.  This scaling of distributions of local quantities
with the value of the global correlation at a given time has been seen
previously in investigations of other models with slow dynamics and no
quenched disorder.\cite{Chamon3} 

\begin{figure}[htb]
\insertpic{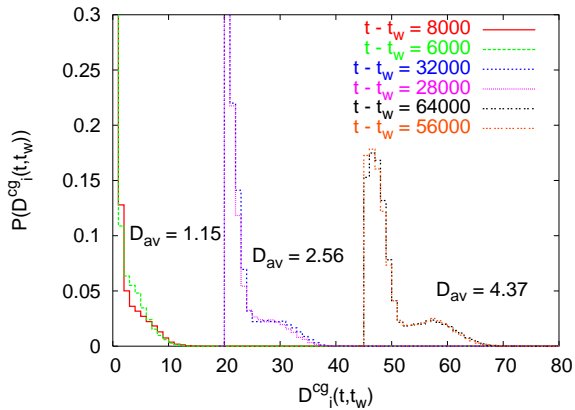}
\caption{(Color online) Distribution of coarse-grained local square displacements for
various values of $t - t_w$, with $x = 0.1$, $T/J = 0.1$,
$J/V = 0.2$, averaged over 20 samples of size $L = 50$.  We show three pairs of
data, which are offset from each other along the $x$-axis, and we show the average
value of the global mean square
displacement $D_{\rm av}$ for each pair.}
\label{fig:dist1}
\end{figure}

The distributions shown in Figs.~\ref{fig:dist1} and \ref{fig:dist2} appear to
be stationary (i.e. depend only on $t - t_w$ rather than $t$ and $t_w$ separately).
Since the distribution is stationary, it
implies that its first moment is also stationary, and hence the distribution itself
scales with the global correlation as follows from the arguments below.
If the correlation and
mean squared displacement are stationary and monotonic, then they can be written
as (using the mean-squared displacement as an example)

\begin{equation}
D(t,t_w) = d(t-t_w)  ,
\end{equation}
and for a given value of $D$ one finds the associated $t-t_w$:
\begin{equation}
D^{-1}[D] = t-t_w  .
\end{equation}
We can use similar arguments for
$C(t,t_w)$.

However, whilst the scaling of the distribution
with the global correlation is trivial, the form of the distribution
is not, and indicates spatially heterogeneous dynamics.
In the distribution of local displacements, 
 it is clear that as $t - t_w$ grows there are two populations of sites,
one with fast dynamics, and another with slow dynamics.  We note that
the distributions displaying two populations of sites have $D(t,t_w) \geq
l_x^2$, implying that one population consists of particles still 
trapped in cages of size $\sim l_x$ and the other is those that have escaped
these cages.  This is
reminiscent of behaviour seen recently in a two dimensional model with
quenched disorder\cite{Kolton} and in a numerical simulation of
colloidal gellation.\cite{Cates}
The local spin correlations do not have a clear separation
into two populations of spins, but there is clearly a wide distribution of local
relaxation times associated with the large spread in correlations.

\section{Stokes-Einstein relation}
\label{sec:SE}

The model studied here has several features that are reminiscent of a
glass-forming liquid.  At high temperatures, the mean-square
displacements are linear in $t$, showing diffusive behaviour, and the
relaxation times associated with this diffusion grow very quickly with decreasing temperature.  
These diffusive properties are what one would expect for a liquid of 
hard-core particles, and what we study is essentially a lattice model of
this type of system.  If the hard-core particles were also allowed to 
have a rotational degree of freedom, then again, the spin of the particles
may be regarded as modelling this physics.  These similarities between
the model we study, and a lattice model for a liquid of hard-core particles, 
inspires us to ask whether well-known properties of liquids are shared by
the model.  In particular, it is interesting to ask whether relationships
similar to the Stokes-Einstein (SE) and Debye-Stokes-Einstein (DSE) relations
hold for translational and spin correlations respectively.

The SE and DSE relations
describe the translation and rotational motion of a {\it large
spherical} particle of radius $R$ in a hydrodynamic continuum in
equilibrium at temperature $T$ with viscosity $\eta$.  The SE relation
predicts the dependence of the {\it translational diffusion coefficient},
${\mathcal D}$, on $T$, $R$ and $\eta$:
\begin{equation}
{\mathcal D} = \frac{k_BT}{6\pi R\eta}
\; .
\label{SE}
\end{equation}
${\mathcal D}$ is
defined from the long-time limit of the mean-square 
displacement, $\langle (\vec x(t)-\langle \vec x(t')\rangle )^2 \rangle =
2 d {\mathcal D} (t-t')$, $d$ is the space dimension, and 
$k_B$ is Boltzmann constant.
Similarly, the DSE relation
predicts the dependence of the {\it rotational correlation time}, $t_{\rm rot}$
on $T$, $R$ and $\eta$:
\begin{equation}
t_{\rm rot}= \frac{4 \pi \eta R^3}{3 k_B T}
\; ,
\label{DSE}
\end{equation}
with $t_{\rm rot}$ extracted from the decay of, say, $\frac{1}{n} \langle
\vec s(t) \cdot \vec s(t')\rangle$, where $n$ is the dimension of the
orientation degree of freedom $\vec s$.  Equations (\ref{SE}) and
(\ref{DSE}) imply that the product ${\mathcal D}t_{\rm rot}$ should not depend on
$T$ and $\eta$.  Even though these relations are derived for a
spherical tracer, in normal liquids they are often satisfied for the
translational and rotational motion of generic probes and the
constituents themselves within a factor of 2.

In a supercooled {\it fragile} liquid the situation changes. While the
rotational motion is consistent with Eq.~(\ref{DSE}), the diffusion of
small probe molecules,~\cite{Cicerone0,Cicerone} as well as the self
diffusion,~\cite{Thurau,Ediger} is much faster than would be expected from 
the 
viscosity dictated by Eq.~(\ref{SE}). The $T$ dependence of ${\mathcal D}$
is not given by $T/\eta$ but there is a ``translational enhancement''
meaning that, on average, probes translate further and further per
rotational correlation time as $T_g$ is approached from above.  For
example, rotational motion in OTP is in agreement with Eq.~(\ref{DSE})
over 12 decades in viscosity while the deviation in the translational
diffusion of {\it small} probe molecules of a variety of shapes, and
the OTP molecules themselves, is such that the product ${\mathcal D}t_{\rm rot}$
increases by up to four orders of magnitude close to $T_g$.  These
results imply that molecules translate without rotating much more than
DSE-SE relations would require.

The mismatch between the translational and rotational motion has been
observed numerically~\cite{SE-num,Harrowell2,Mimo} and it can be described assuming
the existence of dynamic low viscosity
regions,~\cite{Stillinger,Kivelson,Cicerone} a free-energy landscape
based model,~\cite{Dieze} within the random first order transition
scenario,~\cite{Xia-Wolynes} and kinetically facilitated spin models.~\cite{KF,Pan}

\begin{figure}[htb]
\insertpic{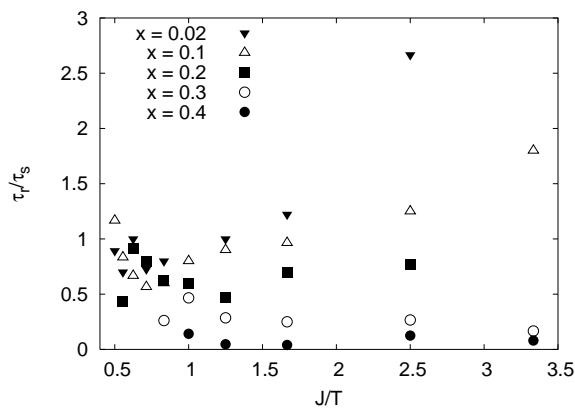}
\caption{Ratio of diffusion and spin relaxation times
as a function of
temperature at $x = 0.02$, 0.1, 0.2, 0.3, and 0.4, with $J/V = 0.2$.}
\label{fig:taucomp}
\end{figure}

Our results in Sec.~\ref{sec:two-time}, where we demonstrate anomalous diffusion in
Figs.~\ref{fig:ddata} and \ref{fig:anomdiff} indicate that the SE relation breaks
down at low temperatures in this model.  To test whether an analogy to the DSE 
holds in this model, in Fig.~\ref{fig:taucomp} we plot the temperature dependence of the
product of the 
translational diffusion coefficient and
the characteristic time for relaxation of the spin correlation 
for five different values of $x$.
We find that that there is strong $x$ dependence
of the ratio $\tau_s/\tau_r \sim {\mathcal D} t_{\rm rot}$ at low temperatures:
for small $x$,
where there is no checkerboard ordering, $\tau_s/\tau_r$ grows at
low temperatures, whereas
the opposite is true for $x \gtrsim 0.2$, where checkerboard ordering is present.

The violation of the DSE relation that we see is much smaller in magnitude
than is generally seen in experiments, but this is likely to be because
the model we investigate displays strong glassy behaviour, whereas 
experiments that study the breakdown of the SE and DSE relations are usually
for fragile glass formers.
The existence of an extra parameter, the doping $x$, also allows for a richer
set of behaviour of ${\mathcal D}t_{\rm rot}$ as a function of 
temperature than is present in supercooled 
liquids.  

\section{Discussion}
\label{sec:disc}

We have studied a classical model for a doped antiferromagnet and
found signatures of slow and heterogeneous dynamics.  At low
temperatures there is anomalous diffusion at lengthscales less than
$l_x = 1/2\sqrt{x}$ and stretched exponential spin relaxations.  We
characterize the temperature below which the slow spin relaxations
occur as $T_{ne}$.  For the doping levels we considered $T_{ne}$ is 
less than the higher of the N\'eel temperature $T_N$ and the
temperature $T_{cb}$, below which checkerboard order is observed.  The
timescales associated with both diffusion and spin relaxation diverge
at low temperatures in an Arrhenius fashion, at all doping levels
considered, behaviour reminiscent of a strong glass former.  
Interestingly, we found that we were able to fit the temperature 
dependence over the entire temperature range with a sub-Arrhenius 
form for both $\tau_r$ and $\tau_s$.
When we investigated the changes in spin configurations
between two different times at several different doping levels, we
found that at low temperatures, the dynamics were spatially
heterogeneous, with regions of high and low mobility -- we also
found that spin flips were most likely to occur adjacent to regions of
high mobility.  This spatially heterogeneous dynamics also manifests itself
in the distributions of local diffusion and spin relaxation.  The distributions of local
square displacements indicate that there are {\it fast} and {\it slow} populations
of particles, corresponding respectively, to those which have escaped, and those which 
are trapped in cages with size of order $l_x$.  The distributions 
of local spin correlations also indicate that there are fast and slow sites, through the
wide range of relaxation times implied by the distribution.
Finally, we found that when
we compare $\tau_s$ and $\tau_r$ there are violations of the Debye-Stokes-Einstein
relation that vary as a function of doping.

On a first glance at Eq.~(\ref{eq:modeldef}) it might appear strange
that there are slow dynamics associated with this model.  There is no quenched
disorder, and there is no explicit frustration in the interactions as
one might expect to see in a glassy model.  
The frustration that leads to slow dynamics appears to be
hidden in the interplay of antiferromagnetic interactions and the
restrictions on particle motion imposed by the level of doping.  The clearest
example of this is the large energy barrier to the motion of a hole to the 
opposite side of a plaquette.  One
way to relate this model to others that have been studied previously
is to think about integrating out the holes, which generates a spin
model with plaquette interactions, which is a generalized version of
the gonihedral models that have been studied and found to have
suggestions of metastable states and glassy dynamics.\cite{Gonihedral}

Whilst we believe that the study of the model in this paper is of
interest in itself, we also note that there is a famous class of doped
two dimensional antiferromagnets, namely the high $T_c$
superconducting cuprates to which this work may have some connections.  There
are some very clear differences between our model and generic models
for these materials, specifically that it is classical rather than quantum
(and hence cannot show superconductivity), and also has Ising spins
rather than Heisenberg spins, leading to an enhancement of long-range
order in two dimensions.  However, the existence of glassy spin
and diffusive dynamics, in the absence of quenched disorder, along
with checkerboard order, even in a
classical model is very interesting.  It also suggests that it might
be interesting to compare quantities analogous to the SE and DSE
relations for spin and charge relaxations in cuprates. 
We intend to investigate this
model with quantum dynamics in future work.  This will hopefully
complement understanding gained in studying spin dynamics in doped two
dimensional quantum antiferromagnets with quenched
disorder.\cite{Chamon2,Yu}

The model displays many of the features that one would expect in a
lattice model for a structural glass (if one imagines the spin as
corresponding to an orientation for a rod-like molecule).  The 
temperature dependence of the relaxation times at low temperatures
is consistent with that of a strong glass-former, although it is
interesting that the temperature dependence over the entire 
temperature range can be fit with a sub-Arrhenius dependence.
It maybe that
dimensionality plays some role in the behaviour that we observe -- it
would be interesting to study this model in three dimensions to see
whether similar behaviour is evident there.  We note that recent
experiments on two dimensional glassy films do show relaxation times that appear
to have sub-Arrhenius behaviour if one fits over a temperature range
that straddles $T_g$.~\cite{Cecchetto}

Another direction of research that we believe may be fruitful is, having
established the existence of slow dynamics in this model to construct
a field theoretic description, and to compare whether the symmetries
found in the case of quenched disorder in the Edwards-Anderson model
\cite{Chamon} are present, and what differences exist between the two
cases.

\section{Acknowledgements}
We thank the anonymous referee for a careful reading of the manuscript.
M.P.K. acknowledges stimulating conversations with Ludovic Berthier,
Horacio Castillo, David Dean, \u{S}imon Kos, Vadim Oganesyan, and
Christos Panagopoulos.  M.P.K. also acknowledges hospitality of
B.U. and a Royal Society Travel Grant at the commencement of this
work. L.F.C. is a member of the Institut Universitaire de France, and this
 research was supported in part by NSF grants DMR-0305482 and
INT-0128922 (C.C.), by EPSRC grant GR/S61263/01
(M.P.K.), an NSF-CNRS collaboration, an ACI-France ``Algorithmes
d'optimisation et systemes desordonnes quantiques'', and the STIPCO European
Community Network (L.F.C.).


\begin{thebibliography}{99}
\bibitem{footnote1} Whilst we set particle moves and spin flips to have equal
     weight, changing the weight of one or the other does not appear to affect
     our results qualitatively.
\bibitem{footnote2} An alternative dynamics that one might consider would
allow the particle to attempt to move to any of its neighbours, where attempts to move
to occupied sites are unsuccessful, but contribute to the number of Monte Carlo steps.  We checked
that this gives the same qualitative features as our results, except with 
longer time-scales.  For $x =0.1$, we find the 
 same value of $T_{ne}$ (defined in Sec.~\ref{sec:two-time}) as with the dynamics we 
use in the rest of the paper.
\bibitem{footnote} Note that in the time window considered, $M_s(t)$ and $M_s(t_w)$ 
saturate at high temperature, but do not for lower temperatures, which is why we 
keep the time depdendence in the staggered magnetization.
\end{thebibliography}
\end{document}